# Tomographic imaging of microvasculature with a purpose-designed, polymeric X-ray contrast agent

Willy Kuo*[a,b], Ngoc An Le[c], Bernhard Spingler[c], Georg Schulz[d], Bert Müller[d],
Vartan Kurtcuoglu*[a,b]

[a]Institute of Physiology, University of Zurich, Winterthurerstrasse 190, 8057 Zurich, Switzerland;
[b]National Centre of Competence in Research, Kidney.CH, Winterthurerstrasse 190, 8057 Zurich, Switzerland; [c]Department of Chemistry, University of Zurich, Winterthurerstrasse 190, 8057 Zurich, Switzerland; [d]Biomaterials Science Center, Department of Biomedical Engineering, University of Basel, Gewerbestrasse 14, 4123 Allschwil, Switzerland

## ABSTRACT

Imaging of microvasculature is primarily performed with X-ray contrast agents, owing to the wide availability of absorption-contrast laboratory source µCT compared to phase contrast capable devices. Standard commercial contrast agents used in angiography are not suitable for high-resolution imaging *ex vivo*, however, as they are small molecular compounds capable of diffusing through blood vessel walls within minutes. Large nanoparticle-based blood pool contrast agents on the other hand exhibit problems with aggregation, resulting in clogging in the smallest blood vessels. Injection with solidifying plastic resins has, therefore, remained the gold standard for microvascular imaging, despite the considerable amount of training and optimization needed to properly perfuse the viscous compounds. Even with optimization, frequent gas and water inclusions commonly result in interrupted vessel segments. This lack of suitable compounds has led us to develop the polymeric, cross-linkable X-ray contrast agent XlinCA. As a water-soluble organic molecule, aggregation and inclusions are inherently avoided. High molecular weight allows it to be retained even in the highly fenestrated vasculature of the kidney filtration system. It can be covalently crosslinked using the same aldehydes used in tissue fixation protocols, leading to stable and permanent contrast. These properties allowed us to image whole mice and individual organs in 6 to 12-month-old C57BL/6J mice without requiring lengthy optimizations of injection rates and pressures, while at the same time achieving greatly improved filling of the vasculature compared to resin-based vascular casting. This work aims at illuminating the rationales, processes and challenges involved in creating this recently developed contrast agent.

## 1. INTRODUCTION

Quantitative assessments of the vascular structure provide valuable insights into various physiological and pathophysiological factors such as vascular growth, tumor development, tissue damage and oxygen transport. Two-dimensional histological sections only capture small sub-volumes, requiring strict adherence to stereological rules[1] to ensure unbiased sampling and prevent mischaracterization of vascular phenotypes with inhomogeneous distribution throughout a tissue. X-ray computed microtomography (µCT) excels in this task by providing fully three-dimensional, micrometer-resolution imaging with large field of view, allowing imaging of whole organs with isotropic image quality. The comparably lower sensitivity of laboratory-source absorption contrast µCT, however, necessitates the use of contrast agents to distinguish blood vessels from the surrounding soft tissues. Phase contrast capable devices represent only a small subset of the currently installed µCT scanners, and are not available to a large part of the scientific community.

The vast majority of commercially available contrast agents are primarily designed, manufactured and used for clinical angiography. Due to a lack of awareness of alternatives, they are often assumed to be the only compounds capable of providing X-ray absorption contrast. As a result, they are frequently employed in other applications which they were not designed for, giving poor results when their chemical and physical properties do not match the needs of the application.

*willy.kuo@uzh.ch, vartan.kurtcuoglu@uzh.ch, https://interfacegroup.ch/

In this work, we present an in-depth discussion of the specific requirements for capillary-resolution imaging of microvasculature, which differ considerably from the requirements for low-resolution, *in vivo* clinical angiography. We further describe how XlinCA, a cross-linkable, polymeric contrast agent was specifically developed to fulfill these requirements,[2] providing insights into the thought processes and rationales used to achieve this goal. Through this, we aim to raise awareness of synthetic chemistry as a versatile tool to create alternatives to commercially available contrast agents, and wish to provide the research community with a basic understanding of how to develop better-suited, purpose-designed contrast agents for their own research questions.

## 2. CHALLENGES AND DESIGN CONSIDERATIONS

### 2.1 Sufficient electron density for laboratory-source absorption contrast

Radiopacity of a contrast agent is primarily dependent on electron density, which is increased by including atomic elements with high atomic number $Z$ such as barium, iodine, gadolinium or lead.[3] This strategy is more flexible compared to using compounds with high intrinsic volumetric mass density, as this is influenced by the surface properties of the compound and cannot be independently adjusted easily. High $Z$ elements confer their radiopacity without requiring any chemical interactions, allowing large leeway in tuning aspects such as molecular weight, hydrophobicity, electric charge or binding sites.

The concentration of high $Z$ elements necessary for microvascular imaging can be very high, however. As an example from our own work, imaging blood vessels with 4 μm voxel size with a General Electric Phoenix nanotom m requires around 80 mg iodine/ml inside the vasculature in order to image a 1 cm$^3$ large volume with sufficient contrast for segmentation.[4] This is several orders of magnitude higher than contrast agents for magnetic resonance imaging (around 0.5 mg gadolinium/ml)[5] or fluorescence microscopy (< 1 μg antibody/ml)[6].

This high concentration requirement has several consequences for contrast agent design: Since high concentrations lead to high osmolarity and viscosity, the fraction of high $Z$ elements within a contrast agent should be maximized to provide the desired contrast at the lowest concentration possible. In addition, the contrast agent needs to be manufactured at much larger scale than fluorescence markers or MRI contrast agents, meaning that the costs per gram have to be kept low.

### 2.2 Hydrodynamic diameter larger than 6 nm, to prevent leakage in permeable vasculature

Angiography contrast agents are low molecular weight compounds capable of passing through the vascular wall within minutes, requiring a bolus injection and immediate imaging in clinical practice. They are also cleared quickly from the blood stream via the kidneys, which is a desirable property in this application to avoid potentially toxic contrast agent accumulation in a patient's body. These characteristics make them unsuitable for organ-scale *ex vivo* μCT imaging with micrometer-resolution, however, due to the longer scan times on the scale of hours and the logistics of extracting and mounting an organ for imaging.

Renal clearance is based on the pore size in the glomeruli, the renal filtration units of the kidney. To stay within the blood stream, a compound needs to feature a hydrodynamic diameter of 6 nm or more[7], which is the equivalent of 65 kDa blood serum proteins such as hemoglobin and albumin. If a contrast agent does not pass this size threshold, it will be filtered through the glomeruli as primary urine into the renal tubules (Figure 1).

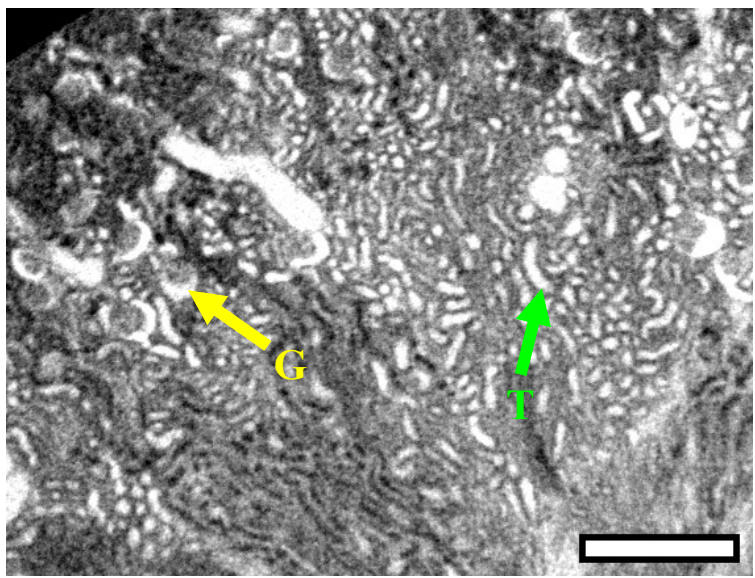

Figure 1. Virtual µCT section of a kidney perfused with polymeric contrast agent that did not have the required minimum molecular weight to avoid glomerular filtration (Compound **5** of Le et al.[2]). As a result, contrast agent can be observed the Bowman's space of the glomerular capsule (G) and inside tubular lumina (T). Voxel size: 4.4 µm, scale bar: 500 µm

### 2.3  Highly water-soluble, to prevent formation of gaps or aggregates

Nanoparticle-based blood pool contrast agents large enough to avoid glomerular filtration are commercially available for preclinical *in vivo* applications. In *ex vivo* applications without active blood flow, however, they tend to aggregate and sediment, clogging small capillaries.[8] The current gold standard in this application is therefore still vascular casting, a method wherein plastic resin mixtures are injected into the vasculature and left to solidify. Radiopacity is provided by additives such as lead chromate microparticles (Microfil, FlowTech Inc.)[9] or iodinated fatty acids (µAngiofil, Fumedica AG)[10]. As these plastic resin mixtures are hydrophobic, they do not pass the hydrated blood vessel walls or glomeruli, resulting in permanent retention within the vasculature.

As the plastic resins and water do not mix, inclusions of residual water within the plastic often result in interrupted blood vessels or formation of resin bubble suspensions (Figure 2). To avoid this behavior, all residual water has to be pushed out of the blood vessels, which requires perfusion with large volume of resin. This can only be achieved with highly optimized perfusion techniques that divert all flow to the organ of interest, which is achieved by closing any other blood flow pathway via ligations.[11] Without these measures, the resin may flow according to the path of least resistance instead, which may circumvent the organ of interest.

Flow rate and thus perfusion pressure is another factor that is difficult to optimize. Since the plastic resins slowly solidify during the procedure, there is a time limit for perfusion the necessary resin volume for flushing out sufficient residual water. The viscosity of the mixture furthermore continuously increases during the procedure, making it difficult to assess and effectively predict the flow rate. This combination makes choosing the correct perfusion pressure a difficult task, as a fine balance has to be found between the maximum perfusion pressure limit for avoiding distension of the blood vessels and the minimum pressure required for flushing out residual water.

All these factors make vascular casting difficult, requiring considerable experience to minimize the number of insufficiently filled blood vessels. As these artefacts are caused by water inclusions, water-soluble compounds avoid these issues entirely.

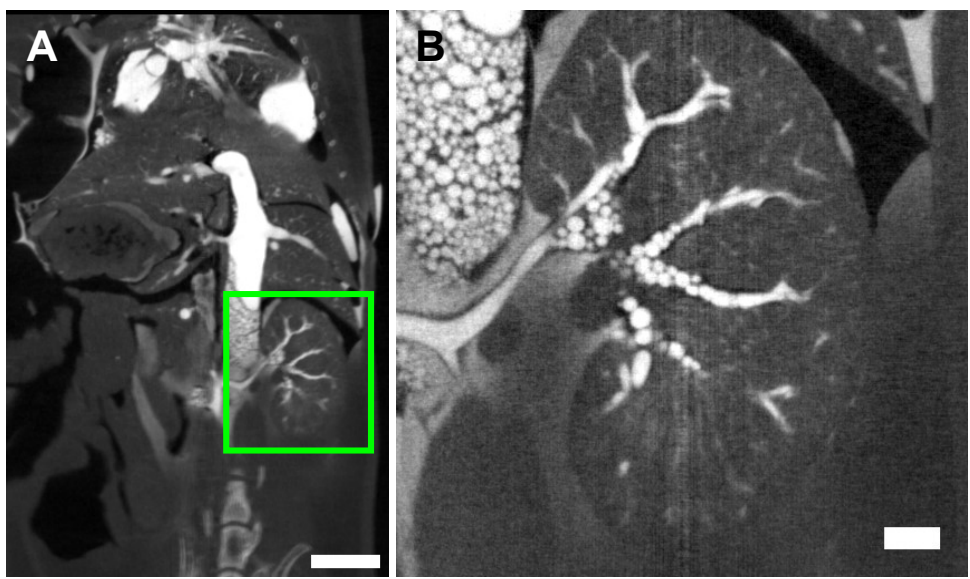

Figure 2. A: Virtual μCT section of the thorax and abdomen of a mouse perfused with a radiopaque plastic resin mixture.[12] Voxel size: 80 μm, scale bar: 5 mm B: Magnified view of the region highlighted in the green box, containing kidney. Venous blood vessels are not continuously filled with plastic resin, instead showing resin bubbles and gaps resulting from residual water included the resin. Voxel size: 20 μm, scale bar: 1 mm.

## 2.4 Cross-linkable, to prevent loss of contrast over time

While water-soluble contrast agents do not pose problems related to homogeneous distribution, this property can also be disadvantageous: as long as the tissue sample is embedded in a water-based medium, diffusion of the contrast agent out of the vasculature can proceed. This not only leads to a reduction in signal inside the vasculature, but also to an increase of signal in the background, leading to substantially reduced contrast over time.

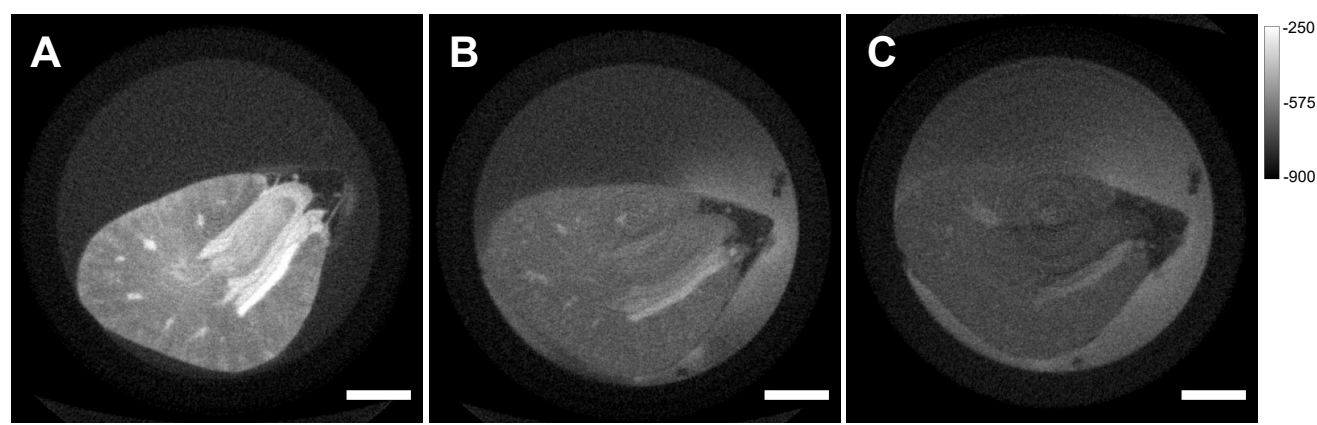

Figure 3. Virtual μCT sections showing the vasculature of mouse kidney perfused with 80 mg iodine/ml of a polymeric, but not fixable contrast agent (Compound **3** of Le et al.[2]). Contrast agent was co-injected with 4.8 % gelatin, and the whole kidney was then embedded in gelatin to restrict leakage of the contrast agent over time. Nevertheless, macroscopic leakage from the kidney vasculature into the surrounding gelatin can be observed. A: Kidney immediately after perfusion. B: Same kidney after three days. C: Section after 48 days. Voxel size: 20 μm, scale bars: 1 mm, gray scale: arbitrary units.

In high-resolution microvascular imaging, this phenomenon is more pronounced due to the small volume, high surface-to-volume ratio and short diffusion distances of capillaries. This means that even ostensibly small reductions in contrast seen on a macroscopic scale as evident in Figure 3 lead to insufficient contrast-to-noise ratio for segmenting capillaries. Using contrast agents with large molecular weight and adding gelling agent such as gelatin or agarose can slow this process, but does not prevent it. Such samples still have to be imaged typically within a day, which can be a considerable logistical challenge depending on the distance to the μCT system. Non-stable samples furthermore cannot be used for imaging with extensive scanning times or rescanned at a later point. Permanent retention of the contrast agent via cross-linking is therefore required both for improving signal-to-noise and for practical use.

## 3. PURPOSE-DESIGNED, POLYMERIC X-RAY CONTRAST AGENT XLINCA

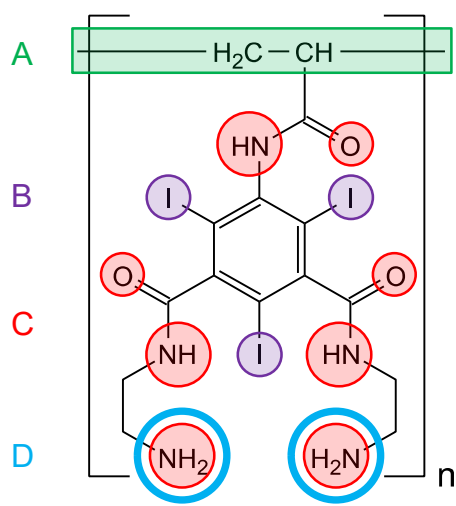

Figure 4. Chemical formula of the repeating unit of the polymeric contrast agent XlinCA. The structural units responsible for the chemical properties have been highlighted with colors. A, Green: The polymer backbone allows the contrast agent to reach high molecular weights, preventing leakage through blood vessel walls. B, Purple: Iodine atoms confer the required radiopacity. C, Red: Hydrophilic groups improve the contrast agent's solubility in water. D, Blue: Primary amine groups allow the polymer to be cross-linked into a hydrogel via aldehyde fixation.

Based on these requirements, we developed XlinCA, a water-soluble, polymeric, cross-linkable contrast agent (Figure 4).[2] We used 5-amino-2,4,6-triiodobenzene-1,3-dicarboxylic acid as starting material, a common precursor molecule for commercial and custom-designed X-ray contrast agents.[13] This starting material features three iodine atoms, representing 68 % of its molecular weight, is amenable to further chemical modifications and can be purchased at low cost.

For increasing the molecular weight, we opted for a linear polymer design, where each repeating unit would carry three iodine atoms derived from the starting material, leading to a fraction of 50 % iodine (w/w) in the final polymer.[2] Typical strategies to increase molecular weight of contrast agents, such as adding polyethylene glycol,[14] are inefficient in that regard, since the increase in size is not linked to a proportional increase in high $Z$ elements. Such a contrast agent would require a much higher concentration to achieve a given concentration of iodine, which in combination with their polymeric nature would result in much higher viscosity, rendering it difficult to inject into microvasculature.

For permanent retention of the contrast agent within the vasculature, we chose to leverage the same aldehyde fixation used in tissue preservation. This would ensure full compatibility with biological tissue preparation and long-term sample storage. The presence of primary amine groups on the contrast agent is sufficient for these purposes, allowing XlinCA to be cross-linked into a hydrogel within minutes by addition of glutaraldehyde.

High water-solubility is conferred by the presence of hydrophilic groups around the monomer. The majority of them are part of the amide linkages (-CONH-) used to add the required chemical structures for the polymerization and the primary amine groups (-NH$_2$), which serve a second purpose beyond facilitating aldehyde fixation by also conferring a large part of the water-solubility.

## 4. METHODOLOGY

### 4.1 Mouse kidney with contrast agent below the glomerular filtration size threshold

The kidneys of a C57BL/6J mouse were perfused via the abdominal aorta with 96 mg iodine/ml of a precursor molecule of the polymeric contrast agent XlinCA without precross-linking (Compound **5** of Le et al.[2]). The expected molecular weight was 20 kDa, below the minimum molecular weight of 65 kDa required to avoid glomerular filtration (Figure 1). The perfusion surgery was performed as described in Kuo et al.[4]. Kidneys were mounted in 1 % agarose in 1.5 ml microcentrifugation tubes and μCT images were acquired with a General Electric Phoenix nanotom m, using an acceleration voltage of 60 kV, a current of 310 μA and 4.4 μm pixel size. Three frames with 0.5 s exposure time were averaged for each of the 1440 projections, resulting in a final scan time of approximately 3 hours per kidney. Reconstruction was performed with the manufacturer's GE phoenix datos|x software. Projections were filtered using a median filter with 3 × 3 pixel kernel size prior to reconstruction. Centers of rotation were determined manually by reconstructing single slices with a series of values.

### 4.2 Whole mouse with radiopaque plastic resin

A whole NMRI mouse was perfused via the left heart ventricle with 20 ml of a heparin solution in phosphate buffered saline, 50 ml 4 % formaldehyde in phosphate buffered saline for fixation and a mixture of the vascular casting resin PU4ii with 107 mg iodine/ml of 1,3-diiodobenzene. The whole mouse was imaged using a PerkinElmer Quantum FX *in vivo* μCT with an acceleration voltage of 50 kV, a current of 200 μA, 80 μm pixel size and a scan time of 3 minutes (Figure 2A). A local tomography scan centered on the kidney was acquired with a pixel size of 20 μm (Figure 2B). Reconstructions were performed automatically by the manufacturer's acquisition software.

### 4.3 Longitudinal study of a mouse kidney with non-fixable contrast agent

The kidneys of a C57BL/6J mouse were perfused via the abdominal aorta with a mixture of 4.8 % gelatin and 80 mg iodine/ml of a precursor molecule of the polymeric contrast agent XlinCA (Compound **3** of Le et al.[2]) as described in Kuo et al.[4]. Kidneys were mounted in 4.8 % gelatin in 1.5 ml and imaged using a PerkinElmer Quantum FX *in vivo* μCT with an acceleration voltage of 70 kV, a current of 200 μA, 20 μm pixel size and a scan time of 3 minutes (Figure 3). Image acquisitions were repeated using the same settings after 3 and 48 days. Reconstructions were performed automatically by the manufacturer's acquisition software.

### 4.4 Mouse kidney with purpose-designed, polymeric x-ray contrast agent XlinCA

The kidneys of a C57BL/6J mouse were perfused via the abdominal aorta with 85 mg iodine/ml of the polymeric, cross-linkable contrast agent XlinCA.[2] The perfusion surgery was performed as described in Kuo et al.[4]. Kidneys were mounted in 1% agar in 0.5 ml microcentrifugation tubes and μCT images were acquired with a General Electric Phoenix nanotom m, using an acceleration voltage of 60 kV, a current of 310 μA and 3.3 μm pixel size (Figure 5). Twelve frames with 0.5 s exposure time were averaged for each of the 1440 projections, resulting in a final scan time of approximately 10 hours per kidney. Reconstruction was performed with the manufacturer's GE phoenix datos|x software. Projections were filtered using a median filter with 3 × 3 pixel kernel size prior to reconstruction. Centers of rotation were determined manually by reconstructing single slices with a series of values.

### 4.5 Whole mouse with purpose-designed, polymeric x-ray contrast agent XlinCA

A whole C57BL/6J mouse was perfused via the left heart ventricle with 100 mg iodine/ml of the polymeric, cross-linkable contrast agent XlinCA, as described in Le et al.[2]. The whole mouse was acquired with a General Electric Phoenix nanotom m, using an acceleration voltage of 60 kV, a current of 310 μA and 20 μm pixel size (Figure 6). Three frames with 0.5 s exposure time were averaged for each of the 1440 projections, resulting in a final scan time of approximately 10 hours per kidney. Reconstruction was performed with the manufacturer's GE phoenix datos|x software. Projections were filtered using a median filter with 3 × 3 pixel kernel size prior to reconstruction. Centers of rotation were determined manually by reconstructing single slices with different values.

# 5. RESULTS

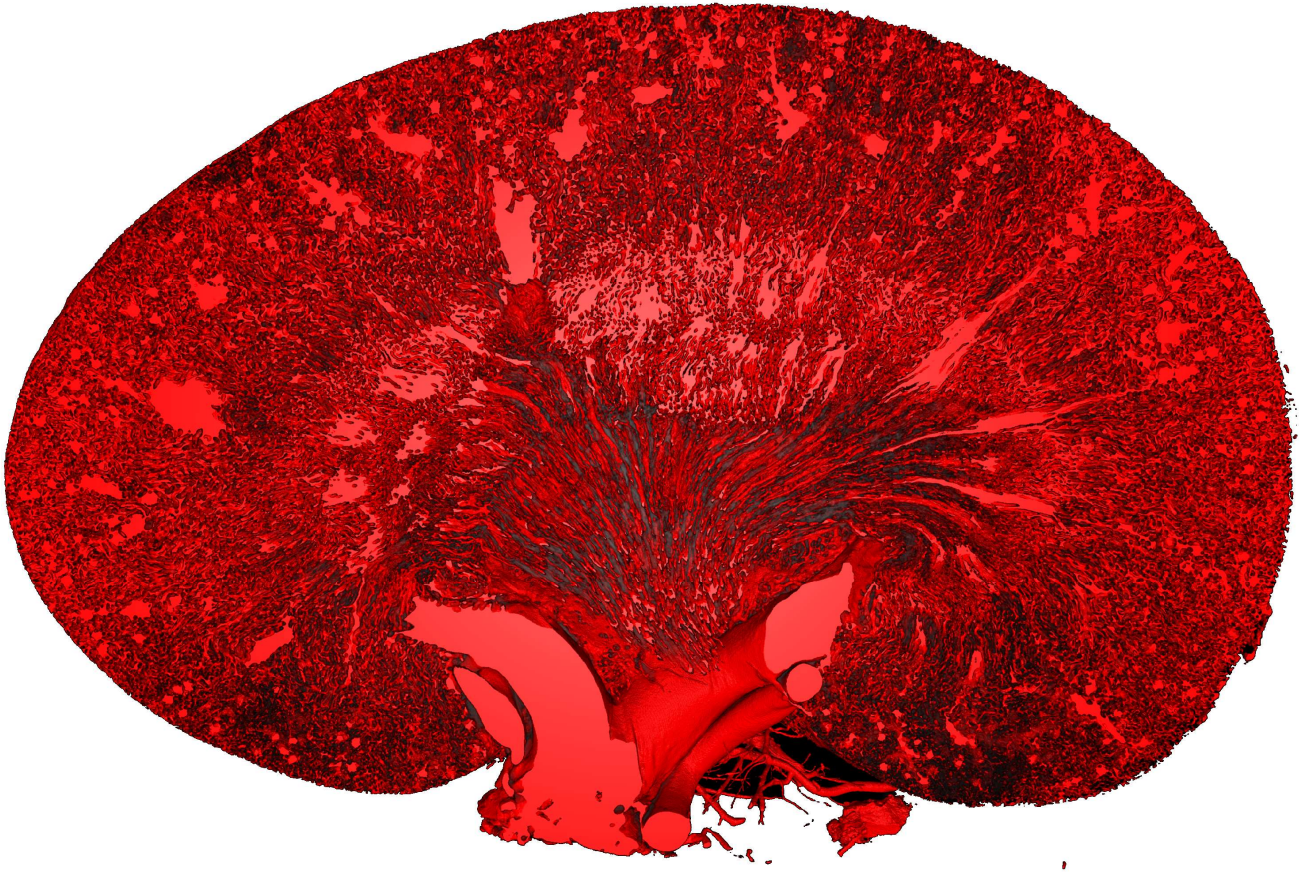

Figure 5. Computer rendering of the segmented vasculature of a mouse kidney perfused via the abdominal aorta with 85 mg iodine/ml of the cross-linkable, polymeric contrast agent XlinCA.[4] All capillaries are visible and no interruptions of the large vessels can be observed, as is expected of a water-soluble compound.

The properties of XlinCA allowed us to image 6 to 12-month-old C57BL/6J mice without requiring lengthy optimizations of injection rates and pressures, while at the same time achieving greatly improved filling of the vasculature compared to resin-based vascular casting (Figures 5 and 6). No leakage of contrast agent into renal tubules could be observed. Ligations of alternate flow pathways are no longer necessary, allowing imaging of the microvasculature of whole bodies and multiple organs of an animal. The covalent cross-linking via aldehydes provides permanent retention of the contrast agent. Samples have been imaged up to a year later with no noticeable reduction in contrast.

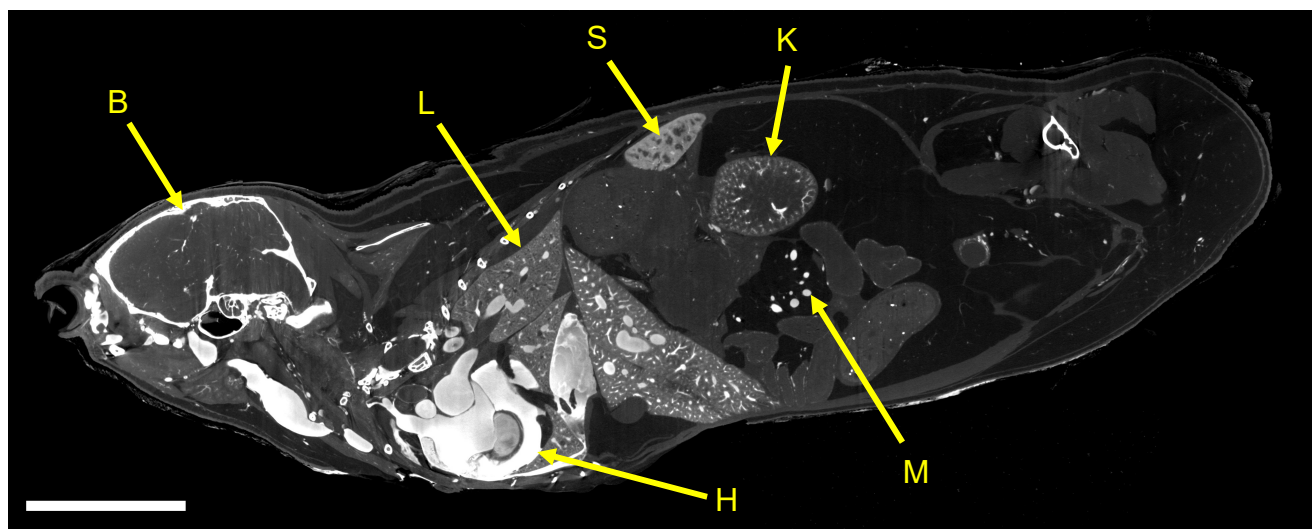

Figure 6. Virtual μCT section of a whole mouse perfused via the left heart ventricle with 100 mg iodine/ml XlinCA.[2] The blood vessels are well visible in multiple organs. B: Brain, L: liver, H: heart, S: spleen, K: kidney, M: mesentery. Voxel size: 20 μm, scale bar: 1 cm.

## 6. DISCUSSION

XlinCA is a contrast agent purpose-designed for *ex vivo* vascular imaging of down to 4 μm wide capillaries. This development was spurred on by the lack of a suitable off-the-shelf solutions. Angiography contrast agents are designed to be cleared from the blood stream *in vivo*, and are thus unsuitable for longer, high-resolution *ex vivo* scans. Nanoparticle-based blood pool contrast agents had the tendency to aggregate when used *ex vivo*, prevented filling of capillaries downstream of the blockages.[8] Suspensions of insoluble salts or microparticles such as barium sulfate show more severe blockages, due to the larger particle size and lack of surface functionalization.[15] Commercially available polymer-based blood pool contrast agents cannot be cross-linked, leaking out of the vasculature over time. Vascular casting with plastic resins is the current gold standard, but the method was historically developed for imaging macroscopic blood vessels and requires considerable expertise and optimization to fill microscopic capillaries reliably.[11,16] Injection with carbon dioxide also could not fill blood vessels with a below 8 μm.[17] This situation necessitated the development of a contrast agent specifically designed to address these issues. The results achieved demonstrate the benefits and advantages of employing a compound well suited to its specific application.

It should be noted that the design considerations described in this work specifically apply for *ex vivo* capillary-resolution vascular imaging. For other applications, different properties would be required. Tissue staining with contrast agents, for example, has the advantage that it does not require distribution of the contrast agent via the vascular system. The method is thus suitable for samples that are difficult or impossible to inject, such as tissue biopsies without intact, enclosed circulatory system. In this approach, the tissue is simply immersed into a contrast agent solution, increasing the X-ray absorption of the tissue, rather than the blood vessel lumina.[18] Surface properties that promote adherence to the tissue of interest are therefore the most important design consideration. For example, positively charged contrast agents have been used to stain negatively charged DNA in cell nuclei[19] or glycosaminoglycans in articular cartilage.[13] Low molecular weight is preferred, as faster diffusion is a benefit, rather than a detriment in this application. Staining protocols are designed for enhancing tissue contrast and not specifically for vascular imaging, however. Blood vessel lumina cannot be distinguished by contrast from other water-filled compartments, such as kidney tubules or brain cerebrospinal fluid spaces. In addition, due to the more limited space inside the tissue, stains may not achieve the same concentration of high Z elements and may require longer scan times or more advanced μCT devices to achieve comparable contrast-to-noise ratios as injectable contrast agents. As an example, a concentration of 2 % iodine (v/v) was reported for staining with iodine potassium iodide,[20] which is considerably lower than the 8 – 10 % iodine (w/v) used for XlinCA or vascular casting.

Further design considerations for a contrast agent can arise if additional features are desired, such as subsequent histological examination. For XlinCA, we have demonstrated compatibility with standard hematoxylin-eosin staining, toluidine blue staining and the periodic acid Schiff reaction, as well as transmission electron microscopy with lead citrate and uranyl acetate contrast enhancement.[4] The glutaraldehyde fixation used for cross-linking may, however, mask antigens required for immunofluorescence or quench fluorescent proteins. To avoid this phenomenon, XlinCA can be cross-linked via formaldehyde, but this requires co-injection with a protein such as bovine serum albumin or a similar compound serving as a bridge, which increases viscosity and osmolarity without a concurrent increase in radiopacity. Another iteration of the contrast agent could be designed to include the required chemical groups for bridging on the molecule itself, reducing the overall concentration of injected compounds.

In conclusion, by having a deeper understanding of the individual chemical and physical properties of contrast agents, we were able to design and produce a polymeric, cross-linkable contrast agent capable of addressing the specific challenges in high-resolution *ex vivo* μCT imaging of microvasculature. Currently commercially available contrast agents could not achieve the same results, as they were designed and optimized for different applications with their own, separate requirements. This highlights the importance of considering compounds beyond the dedicated angiography contrast agents, and the benefits of selecting or developing contrast agents with the specific properties required for imaging experiments.

## ACKNOWLEDGEMENTS

This work was financially supported by the University of Zurich and the Swiss National Science Foundation through NCCR Kidney.CH, R'Equip 133802 and grant 205321_153523.

## CONFLICT OF INTEREST

A patent application for the polymeric contrast agent XlinCA has been filed by the University of Zurich (PCT/EP2020/084771).